\documentclass[12pt]{iopart}

\begin{document}

\title[Gaseous Dark Matter Detectors]{Gaseous Dark Matter Detectors}
\author{G Sciolla}
\address{Massachusetts Institute of Technology, 
77 Massachusetts Avenue, Cambridge, MA 02139, USA}
\ead{sciolla@mit.edu} 

\author{C J Martoff}
\address{Temple University,       
1900 N. 13-th Street, Philadelphia, PA 19122, USA} 
\ead{martoff@temple.edu} 

\begin{abstract}
Dark Matter detectors with directional sensitivity have 
the potential of yielding an unambiguous positive observation of WIMPs 
as well 
as discriminating between galactic Dark Matter halo models.  
In this article, we introduce the motivation for directional detectors, 
discuss the experimental techniques that make directional detection possible, 
 and review the status of the experimental effort in this field.  
\end{abstract}

\pacs{95.35.+d, 14.80.Ly, 29.40.Cs, 29.40.Gx, 29.40.-n} 

\vspace{2pc}
{\noindent{\it Keywords}: Dark matter, Directional detector, Gas detector, WIMP, neutralino,  TPC}
\submitto{\NJP}
\maketitle


\section{Introduction}

Astronomical and cosmological observations have recently shown that 
Dark Matter (DM) 
is responsible for 23\% of the energy budget of the Universe and 83\% of its mass~\cite{Hinshaw2008}. 
The most promising candidate for Dark Matter is the so-called Weakly Interacting Massive 
Particle (WIMP).  The existence of WIMPs is independently suggested by
considerations of Big Bang cosmology and theoretical supersymmetric particle 
phenomenology~\cite{LeeWeinberg,Weinberg82,Jungman1996}.

Over the years, many direct detection experiments have been performed to search for  
nuclear recoils due to elastic scattering of WIMPs off the nuclei 
in the active volume of the detector. 
The main challenge 
for these experiments is to 
suppress the backgrounds that mimic WIMP-induced nuclear recoils. 
Today's leading experiments have achieved excellent rejection of electromagnetic 
backgrounds, i.e., photons, electrons and alpha particles,  
that have a distinct signature in the detector.  
However, there are sources of background for which the detector response is nearly 
identical to that of a WIMP-induced recoil, such as 
the coherent scattering of neutrinos from the sun~\cite{Monroe2007}, or    
the elastic scattering of neutrons produced either by natural radioactivity 
or by high-energy cosmic rays. 

While neutron and neutrino interactions do not limit  
today's experiments, they are expected to become dangerous sources of background when 
the scale of DM experiments  grows to fiducial masses of several tons.  
In  traditional counting experiments, the presence of such backgrounds 
could  undermine the unambiguous identification of a Dark Matter signal because  
neutrinos are  impossible to suppress by shielding 
and underground neutron backgrounds are notoriously difficult to predict~\cite{Mei2006}. 

An unambiguous positive identification of a Dark Matter signal
even in presence of unknown amounts of irreducible backgrounds 
could still be achieved if one could correlate the observation of 
a nuclear recoil in the detector with some unique astrophysical signature which no 
background could mimic. This is the idea that motivates directional detection of Dark Matter.

\subsection{The Dark Matter Wind }

The observed  rotation curve of our Galaxy suggests that at the galactic radius of the sun 
the galactic potential 
has a significant contribution from Dark Matter. 
The Dark Matter distribution in our Galaxy, however,  is poorly constrained. 
A commonly used DM distribution, the standard dark halo 
model~\cite{SmithLewin1990}, assumes a non-rotating, isothermal sphere extending out to 
50 kpc
from the galactic  center. The DM velocity is described by a Maxwell-Boltzmann 
distribution 
with dispersion $\sigma_v=155$ km/s. 
Concentric with the  DM halo is the galactic disk of 
luminous  ordinary matter, rotating with respect to the halo, with an average 
orbital velocity 
of about 220 km/s at the radius of the solar system. 
Therefore in this model, an observer on Earth would see a wind of DM particles with 
average velocity of 220 km/s. 

The Dark Matter wind creates two observable effects. 
The first was pointed out in 1986 by Drukier, Freese, and Spergel~\cite{Drukier1986} who predicted that 
the Earth's motion relative to the galactic halo leads to an annual modulation 
of the rates of interactions observed above a certain threshold in direct detection experiments. 
In its annual rotation around the sun, the Earth's orbital velocity has a component that 
is  anti-parallel to the DM wind during the summer, 
and parallel to it during the winter. As a result,
the apparent velocity of the DM wind will increase (decrease) by about 10\% in summer (winter), 
leading to a corresponding increase (decrease) of the observed rates in DM detectors. 
Unfortunately, this effect is difficult to detect because the seasonal modulation is expected to be small (a few \%) 
and very hard to disentangle from other systematic effects, such as the seasonal dependence of background rates. 
These experimental difficulties cast a shadow on the recent claimed observation of the yearly asymmetry by the DAMA/LIBRA 
collaboration~\cite{Bernabei2008}. 

A larger modulation of the WIMP signal 
was pointed out by Spergel~\cite{Spergel} in 1988. 
The Earth spins around its axis with a period of 24 sidereal hours. Because its rotation axis is 
oriented at 48$^\circ$  with respect to the direction of the DM wind, an observer on Earth sees the average 
direction of the WIMPs change  
by 96$^\circ$  every 12 sidereal hours.
This modulation in arrival direction should be resolvable 
by a Dark Matter directional detector, e.g., a detector able to determine the direction of the DM particles. 
Most importantly, no known 
background is correlated with the direction of the DM wind. Therefore, a directional  detector could hold the 
key to the unambiguous observation of Dark Matter. 

In addition to background rejection, 
the determination of the direction of the arrival of Dark Matter particles can 
discriminate~\cite{Copi1999,Vergados2003,Morgan2004,Freese2005,Alenazi2008}
between various DM halo distributions including the standard dark halo model,  models with streams of WIMPs, 
the Sikivie late-infall halo model~\cite{Sikivie1999,Tkachev1997,Sikivie1995}, and other anisotropic models.  
The discrimination power is 
further enhanced 
if a determination of the sense as well as the direction of WIMPs is possible~\cite{Green2007}.    
This capability  makes directional detectors 
unique observatories for underground WIMP astronomy.

\subsection{ Directional Dark Matter Detection } 

When Dark Matter particles interact with regular matter, they scatter elastically off the atoms and 
generate nuclear recoils with typical energies $E_R$ of a few tens of keV, as explained in more detail in 
section~\ref{NuclearRecoils}.   
The direction of the recoiling nucleus encodes the direction of the incoming DM particle.  
To  observe the daily modulation in the direction of the DM wind, an  angular resolution  of  20--30 degrees 
in the reconstruction of the recoil nucleus is sufficient, because the intrinsic spread in direction 
of the DM wind is $\approx$ 45 degrees. 
Assuming that sub-millimeter tracking resolution can be achieved, the  length of 
a recoil track has to be of at least  1--2 mm, which can be  
 obtained by using a very dilute gas as a target material. 

An ideal directional detector should  provide a 3-D vector reconstruction of the recoil track with a spatial 
resolution of a few hundred microns in each coordinate, and 
combine a  very low energy threshold with an 
excellent background rejection capability. Such a detector would be able to reject isotropy of the recoil direction, 
and hence identify the signature of a WIMP wind, with just a handful of events~\cite{Morgan2004}. 

More recently, Green and Morgan~\cite{Green2007}  studied how the number of events  necessary to 
detect the WIMP wind depends on the detector performance 
in terms of 
energy threshold,  background rates, 2-D versus 3-D reconstruction of the nuclear 
recoil, and ability to determine the sense of the  
direction by  discriminating between the ``head'' and ``tail'' of the recoil track. 
The default configuration used for this study assumes a CS$_2$ gaseous TPC running at 0.05 bar 
using 200 $\mu$m pixel readout providing 3-D reconstruction of the nuclear recoil 
and  ``head-tail'' discrimination. 
The energy threshold is assumed to be 20 keV, with perfect background rejection. 
In such a configuration, 7 events would be sufficient to establish observation of the 
WIMP wind at 90\% C.L.. 
In presence of background with S/N=1, the number of events necessary to reject isotropy 
would increase by a factor 2. 
If only 2D reconstruction is available, the required number of events doubles compared to 
the default configuration. 
``Head-tail'' discrimination turns out to be the most important capability: if the sense 
cannot be measured, the 
number of events necessary to observe the effects of the WIMP wind increases by one order of magnitude.

\section{Nuclear Recoils in Gaseous Detectors}
\label{NuclearRecoils} 

To optimize the design of gaseous detectors for directional detection of 
Dark Matter one must be able to calculate  the recoil 
atom energy spectrum expected for a range of WIMP parameters 
and halo models. 
The detector response in the relevant energy range must also be predictable. 
The response will be governed first and foremost by the track 
length and  characteristics (multiple scattering)  as a 
function of recoil atom type and energy. Since gas detectors 
 require ionization for detection,
design also requires knowledge of the 
ionization yield in gas and its  distribution along the track 
as a function of recoil atom type and energy, and possibly 
electric field. 

The large momentum 
transfer necessary to produce a detectable recoil in gas implies 
that the scattering atom can be treated as a free particle, making calculations of the 
recoil spectrum essentially 
independent of whether the target is a solid, liquid, or gas.

 An estimate of the maximum Dark Matter recoil energy for simple halo models is given by  the kinematically allowed energy transfer from an infinitely 
heavy halo WIMP with velocity equal to the galactic 
escape speed.  This speed is locally about 500-600 km/sec~\cite{rave}; 
WIMPS with higher velocities than this would not be gravitationally bound in the halo and 
would presumably be rare. 
The corresponding maximum energy transfer amounts  to $<$ 10 keV/nucleon. The integrated rate will be concentrated at lower energies than this, at least in 
halo models such as the isothermal sphere.  For that model, the recoil energy ($E_R$) distribution~\cite{SmithLewin1990}
is proportional to $\exp(-E_R/E_I)$, with $E_I$ a constant that depends on 
the target and WIMP masses and the halo model.  
For a 100 GeV WIMP and the isothermal halo model parameters of Ref.~\cite{SmithLewin1990}, $E_I / A$ 
varies from 1.05 to 0.2 keV/nucleon for target mass numbers from 1 to 131.  These are very low energy particles, well below the Bragg Peak at $\sim$200--800 keV/A.  
In this regime dE/dx {\it decreases} with decreasing energy, and the efficiency of ionization is significantly reduced. 

\subsection{Lindhard Model for Low-Energy Stopping}
The stopping process for such low energy particles in  
homoatomic\footnote{A 
homoatomic molecular entity is a molecular entity consisting of one or more atoms of the same element.} 
 substances was treated by Lindhard, Scharff, and Schiott~\cite{Lindhard1963,Lindhard-int} (LSS).  
This treatment has stood the test of time and experiment, 
making it worthwhile to summarize the results here.  

As is now well-known, the primary energy loss mechanisms for low energy particles in matter can be divided 
into ``nuclear stopping'', due to atom-atom scattering, and ``electronic stopping'', due to atom-electron scattering. 
These mechanisms refer only to the initial interaction causing the incident particle to lose energy.  
Nuclear stopping eventually contributes to electronic excitations and ionization, 
and electronic stopping eventually contributes to thermal excitations~\cite{Lindhard-int}.  

In Ref.~\cite{Lindhard1963} the stopping is described using a Thomas-Fermi atom model to obtain 
numerical results 
for universal stopping-power curves in terms of two variables, the scaled energy  $\epsilon=E_R/E_{TF}$,  
and the scaled range  $\rho=R/R_{TF}$, 
where $E_R$ and $R$ are respectively the energy and the stopping distance of the 
recoil,     
and $E_{TF}$ and $R_{TF}$ are scale factors\footnote{The
 scale factors are (in cgs-Gaussian units): $E_{TF} =
 \frac{e^2}{a} Z_i Z_T \frac{M_i +M_T}{M_T}$, $R_{TF} =
 \frac{1}{4 \pi a^2 N} \frac{(M_i + M_T)^2}{M_i M_T}$.  Here,
 $N$= number density of target atoms, subscripts i and T refer
 to the incident particle and the target substance, and $a = a_0
 \frac{.8853}{\sqrt{Z_i ^{2/3} + Z_T ^{2/3}}} $, with $a_0$ the
 Bohr radius.}.

In Ref.~\cite{Lindhard1963} it was  
shown that nuclear stopping dominates in the energy range where most of the
 rate for Dark Matter detection lies. This can be seen as 
 follows. The scaled variables $\epsilon$ and
 $\rho$ depend algebraically on the atomic numbers and
 mass numbers of the incident and target particles.
 The scale factor $E_{TF}$ corresponds to 0.45
 keV/nucleon for homoatomic recoils in Carbon, 1.7 keV/nucleon
 for Ar in Ar and 6.9  keV/nucleon for Xe in Xe. Nuclear
 stopping $\frac{d\epsilon_n}{d\rho}$ was found to be larger
 than the electronic stopping $\frac{d\epsilon_e}{d\rho}$ for
 $\epsilon < 1.6$, which covers the energy range $0 < E_R <
 E_I$ where 
 most of the Dark Matter recoil rate can be expected.   

Because of the dominance of nuclear stopping, detectors can be expected to respond differently 
to Dark Matter recoils than to radiations such as x-rays or even $\alpha$ particles, for 
which electronic stopping dominates.  
Nuclear stopping yields less ionization and electronic excitation per unit energy loss 
than does electronic stopping, implying that the W factor, defined as the energy loss required 
to create one ionization electron, will be larger for nuclear recoils.  
Reference~\cite{Lindhard-int} presents calculations of the ultimate energy loss partitioning  
between electronic and atomic motion.       
Experimenters use empirical ``quenching factors" to describe the variation of  
energy per unit of ionization (the ``W" parameter) compared to that from x-rays. 

The different microscopic distribution of ionization in tracks dominated by nuclear stopping 
can also lead to unexpected changes in the interactions of ionized and electronically excited 
target atoms (e.g., dimer formation, recombination).
Such interactions are important for particle identification signatures such as the 
quantity and pulse shape
of scintillation light output, the variation of scintillation pulse shape with 
applied electric field, and the field variation of ionization charge collection 
efficiency.  
Such effects are observed in gases~\cite{White2007,Martin2009}, and even more strongly 
in liquid and solid targets~\cite{Aprile2006}. 

Electronic stopping~\cite{Lindhard1963} was found to vary as $\frac{d\epsilon_e}{d\rho} = k \sqrt{\epsilon}$ 
with the parameter $k$ varying only from 0.13 to 0.17 for homonuclear recoils in A=1 to 131\footnote{The 
parameter $k \stackrel{.}{=} \frac{0.0793Z_1^{1/6}}{(Z_1^{2/3} + Z_2^{2/3})^{3/4}} 
\left[\frac{Z_1Z_2(A_1+A_2)^3}{A_1^3A_2}\right] ^{1/2}$ becomes substantially larger only for light  recoils 
in heavy targets.}. Let us define the total stopping as $\frac{d\epsilon}{d\rho}= \frac{d\epsilon_n}{d\rho} + 
\frac{d\epsilon_e}{d\rho}$ and the total scaled range as
$\rho_o = \int _0 ^\epsilon \frac{d\epsilon}{(\frac{d\epsilon}{d\rho})}$. The relatively small contribution of electronic stopping and the small variation in $k$ for homoatomic recoils, makes the total scaled range for this case depend on the target and projectile almost entirely through $E_{TF}$. 

Predictions for the actual range of homoatomic recoils can be obtained from the nearly-universal scaled range curve
as follows. Numerically integrating the stopping curves of Ref.~\cite{Lindhard1963} with $k$ set to 0.15 gives a scaled range curve that fits the surprisingly simple expression 
\begin{equation}
\rho_o \stackrel{.}{=} 2.04 \epsilon + 0.04
\label{eq:range}
\end{equation}
 with accuracy better than 10\% for $0.12 < \epsilon < 10 $. 
According to the formula given earlier, the scale factor $R_{TF}$ lies 
between 1 and 4 $\times$ 10$^{17}$ atoms/cm$^2$ for 
homoatomic recoils in targets with $12 \leq A \leq 131$.  
Thus the model predicts ranges of several times 10$^{17}$ atoms/cm$^2$ at $E_R = E_I$.  
This is of the order of a few mm for a monoatomic gas at 0.05 bar.  
As a consequence, tracking devices for Dark Matter detection must provide accurate 
reconstruction of tracks with typical lengths between 1 and a few mm 
while operating at pressures of a small fraction of an atmosphere.

When comparing LSS predictions with experimental results, two correction
factors must be considered.  First, 
the widely-used program SRIM~\cite{SRIM} produces range-energy tables which contain the ``projected 
range",  while LSS calculate the path length along the track.  On the other hand, many older experiments  
report ``extrapolated ranges", which are closer in magnitude to the path length than to the ``projected 
range".  To compare the SRIM tables with LSS, the projected range should be multiplied 
by a factor~\cite{Lindhard1963} $(1+\frac{M_T}{3M_P})$ where $M_T$ and $M_P$ are the 
target and projectile masses.  This correction has generally been applied in the next section, where
experimental data are discussed.

In addition, it must be noted that the LSS calculations described above were obtained for solids. 
Therefore, one should consider a gas-solid correction in ranges and stopping powers, as  
discussed by Bohr, Lindhard and Dan~\cite{BLD}. In condensed phases, the higher collision frequency 
results in a higher probability for stripping of excited electrons before they can relax, 
which leads to a higher energy loss rate than for gases.  This correction is rather 
uncertain and has generally not been applied in the following section of this paper.

Finally, numerical calculations to extend the LSS model to the case of targets of mixed 
atomic number are given in Ref.~\cite{Hitachi2008}.

\subsection{Experimental Data on Low Energy Stopping in Gases} 

The literature of energy loss and stopping of fast particles in matter is vast and still 
growing~\cite{Ziegler1985,Sigmund1998}.  
However, there is not a lot of experimental 
data available for particle ranges and ionization yields in gas at the very low energies 
typical of Dark Matter recoils, where E/A $\sim$ 1 keV per nucleon.   
Comprehensive collections of citations for all energies are available~\cite{SRIM,MSTAR}, 
upon which the widely-used 
theory-guided-fitting computer programs SRIM and MSTAR~\cite{MSTAR} are based.  
Several 
older references~\cite{Evans1953, Lassen1964, Cano1968}
still appear representative of the available direct measurements at very low energy. 
More recent studies~\cite {SnowdenIfft2003} provide indirect information based 
on large detector  simulations.

Both references~\cite{Evans1953} and~\cite{Lassen1964} used accelerated beams of 
He, N, Ne, Ar and $^{24}$Na, $^{66}$Ga, and $^{198}$Au in differentially pumped gas target chambers filled 
with pure-element gases.  In~\cite{Evans1953} the particles were detected with an ionization chamber, 
 while in~\cite{Lassen1964}  
radioactive beams were used.  The stopped particles were collected on segmented walls of the target chamber
and later counted. 
Typical results were ranges of  2(3.2) $\times$ 10$^{17}$ atoms/cm$^2$ for 26(40) keV Ar$^+$ in Argon.  
The fit to LSS theory given above 
predicts ranges that are shorter than the experimental results by 10-40\%,
 which is consistent with experimental comparisons given by LSS.  
Accuracy of agreement with the prediction from the SRIM code is about the same.  As in all other cases
discussed below, the direction of the deviation from LSS is as expected from the gas-solid effect
mentioned in the previous section.

In Ref.~\cite{SnowdenIfft2003}  nuclear recoils from $^{252}$Cf neutrons were recorded by 
 a Negative Ion Time Projection Chamber (NITPC) filled with 40 Torr CS$_2$.  
   The device was simulated 
fitting the observed pulse height and event size distributions.  
The best fit range curves given for C and S recoils 
in the gas are 10-20\% higher at 25-100 keV than LSS predictions computed by the present authors   
by assuming simple additivity of stopping powers for the constituent atoms of the polyatomic 
gas target.
 
\subsection{Ionization Yields}

Tracking readouts in gas TPC detectors are sensitive only to ionization of the gas.  
As noted above, both nuclear and electronic stopping eventually contribute to both electronic excitations 
(including ionization)  and to kinetic energy of target atoms, as primary and subsequent generations of  
collision products interact further with the medium.  Some guidance useful for design purposes is 
available from Ref.~\cite{Lindhard-int}, where the energy cascade was treated numerically using 
integral equations.  In terms of the scaled energy $\epsilon$ and the electronic stopping 
coefficient $k$ introduced above, the (scaled) energy $\eta$ ultimately transferred to electrons was found to 
be well approximated~\cite{SmithLewin1996} by 
$\eta = \frac{\epsilon}{1+\frac{1}{k\dot g(\epsilon)}}$ with
$g(\epsilon)= \epsilon + 3 \epsilon^{0.15} + 0.7 \epsilon^{0.6}$.
This function interpolates smoothly from $\eta = 0$ at $\epsilon = 0$ to $\eta = \epsilon$ 
for $\epsilon \rightarrow \infty$, giving  $\eta = 0.4$ at $\epsilon = 1$.  
In other words, this theory predicts only about 40\% as much ionization per unit of 
energy deposited by Dark Matter recoils as by low LET radiation such as electrons ejected by x-rays.

Several direct measurements of total ionization by very low energy particles are available in literature. 
Many of these results are for recoil nuclei from alpha decays~\cite{Cano1965,Cano1968,Stone1957}. 
These $\sim$~100 keV, A $\sim$ 200 recoils are of
interest as backgrounds in Dark Matter experiments, but their scaled energy $\epsilon \cong 0.07$ is 
 below the range of interest for most WIMP recoils.
Measured ionization yield parameters W were typically 100-120 eV/ion pair, in good agreement with the approximate formula for $\eta$ given above.
  Data more applicable to Dark Matter recoils are given in Refs.~\cite{Phipps1964,Boring1965,McDonald1969,Price1993}.  
Some representative results from these works include~\cite{Boring1965} 
  W = 91 (65) eV/IP for 25 (100) keV Ar in Ar, both values about 20\% higher than would be predicted by the 
preceding approximate LSS expression.  Higher W for gases than for condensed media is expected~\cite{BLD} as 
mentioned above.  Ref.~\cite{McDonald1969} measured total ionization from particles 
with 1 $<$ Z $<22$ in methane.  While  in principle the LSS treatment does not apply to heteroatomic gases, 
using the LSS prescription to predict the W factor for a carbon target (rather than methane) yields 
a value that is 15\% lower than the experimental results.    
  
The authors of Ref.~\cite{SnowdenIfft2003} also fit their data to derive W-values for C and S recoils.  
Their best-fit values are again 10-25\% higher than an LSS-based estimate by the present author using 
additivity. 

  To summarize, most of the Dark Matter recoils expected from an isothermal galactic halo have very low energies,
and therefore nuclear stopping plays an important role.  The sparse available
   experimental data on track lengths and ionization yields agrees at the $\sim$20\% level with simple 
approximate formulas based on the Lindhard model.
Without applying any gas-phase correction, LSS-based estimates
for range tend to be slightly longer than those experimentally measured in gases.  The predicted ionization parameter W also tends to be slightly lower than the experimental data.
This situation is adequate for initial design of detectors, but with the present literature base, 
each individual experiment will require its own dedicated calibration measurements.

\section{Considerations for Directional Detector Design}       
             
\subsection{Detector Architecture}      

 From the range-energy discussion in the previous section, 
 we infer that 
 track lengths of typical 
 Dark Matter recoils will be only of the order of 0.1 $\mu$m in condensed matter, 
 while  track lengths of up to a few millimeters  are expected in gas 
 at a tenth of the atmospheric pressure.
 Several techniques relevant to direction-sensitive detection using condensed matter 
 targets have been reported, including track-etch analysis of 
ancient mica~\cite{Bander1995}, bolometric detection of surface sputtered 
atoms~\cite{Martoff1996}, and use of nuclear emulsions~\cite{Natsume2007}.  The 
ancient mica etch pit technique was actually used to obtain Dark Matter limits. 
 However, recently the 
 focus of directional Dark Matter detection has  shifted to 
 low-pressure gas targets, and that is the topic of the present review.

 The TPC~\cite{NygrenTPC,Fancher1979} is the natural 
detector architecture for gaseous direction-sensitive Dark Matter 
 detectors, and essentially all experiments use this configuration. The active target 
 volume contains only the active gas, free of background-producing material. Only 
 one wall of the active volume requires a readout system, leading to 
 favorable cost-volume scaling. TPCs with nearly 100 m$^3$ of active volume have 
 been built for high energy physics, showing the possibility of large active 
 masses.

 \subsection{Background Rejection Capabilities } 

 Gaseous DM detectors have excellent background rejection capability for 
 different kinds of backgrounds. 
 First and foremost, direction sensitivity gives gas detectors the capability of statistically rejecting 
neutron and neutrino backgrounds.  
 In addition, tracking also leads 
 to extremely effective discrimination against x-ray and $\gamma$-ray 
 backgrounds~\cite{Snowden-Ifft:PRD2000,Sciolla:2009fb}. The energy loss 
 rates for recoils discussed in the previous section are hundreds of times 
 larger than those of electrons with comparable total energy. The resulting much 
 longer electron tracks are easily identified and rejected in any direction-sensitive detector. 
  Finally, the measured rejection factors for gamma rays vs. nuclear 
 recoils varies  between 10$^4$ and 10$^6$ depending on the experiment~\cite{Miuchi2007-58,SnowdenIfft2003,Dujmic2008-58}.

 \subsection{Choice of Pressure}

It can be shown 
 that there is an optimum pressure for operation of any given direction sensitive 
 WIMP recoil detector.  This optimum pressure depends on 
 the fill gas, the halo parameter set and WIMP mass, and 
 the expected track length threshold for direction measurement. 

The total sensitive mass, and hence the total number 
 of expected events, increases 
proportionally to the product of the pressure $P$ and the active volume $V$. 
Equation~\ref{eq:range} 
 above shows that the range in atoms/cm$^2$ for WIMP recoils is approximately 
 proportional to their energy. Since the corresponding range in cm is inversely 
 proportional to the pressure ($R \propto E_r/P$), the energy threshold imposed 
 by a particular minimum track length $E_{r,min}$ 
will scale down linearly with decreasing 
 pressure, $E_{r,min} \propto R_{min} P$, where $R_{min}$ is the shortest detectable track length. 
For the exponentially falling recoil energy spectrum 
 of the isothermal halo~\cite{SmithLewin1996} the fraction of recoils above a given energy 
 threshold is proportional to $\exp(-E_{min}/E_0 r)$. Hence the rate of tracks longer
than the tracking threshold R$_{min}$ will scale as $N \propto PV \exp(-\xi R_{min}P)$, with $\xi$ 
 a track length factor depending on the target gas, WIMP mass, halo 
 model, etc., and the track length threshold $R_{min}$ depending on the readout 
 technology and the drift distance. This expression has a maximum at $P_{opt} 
 = 1/[\xi R_{min}]$, which shows that the highest event rate is obtained by taking advantage
of improvement in tracking threshold to run at  
higher target pressure. Operating at this optimum pressure, the track-able event rate still scales as $P_{opt}V$, which increases linearly as the tracking threshold 
 decreases. Achieving the shortest possible tracking threshold $R_{min}$ 
 is seen to be the key to sensitive experiments of this type.

 \subsection{Tracking Limit due to Diffusion}        
 Diffusion of track charge during its drift to the readout plane sets 
 the ultimate limit on how short a track can be measured in 
 a TPC. Diffusion in gases has a rich phenomenology for which only 
 a simplified discussion is given here. More complete discussion with references to 
 the literature is given by Rolandi and Blum~\cite{RnB}.     
             
 For  low values of electric fields, elementary kinetic theory arguments predict equal 
 transverse and  longitudinal diffusion  
 to the drift field $E_d$, with the rms diffusion spread  $\delta$ given by 
             
 \begin{equation}            
 \label{eq:diff}            
 \delta = \sqrt{\frac{2kTL}{eE_d}} = 0.7 mm \sqrt{\frac{[L/1m]}{[E_d/1 kV/cm]}}.      
 \end{equation}            
             
 Here $k$ is the Boltzmann constant, $T$ the gas temperature, and $L$ the drift 
 distance. 
No pressure or gas  dependence appears in this equation. The diffusion decreases inversely as the square 
 root of the applied drift field. Increasing the drift field would appear 
 to allow diffusion to be reduced as much as desired, allowing large 
 detectors to be built while preserving good tracking resolution.    
             
 However, in reality diffusion is not so easily controlled. 
 The low-field approximation 
 given by Equation~\ref{eq:diff} holds only below a certain 
  maximum drift field value $E_d^{max}$, which depends on 
 the pressure and target gas.
 The drift field must not violate 
 the condition $eE_d^{max} \lambda << kT$, where the effective mean free path 
 $\lambda = 1/f n \sigma$ decreases inversely as the pressure. Here $\sigma$ 
 is the average total  cross section for scattering of the drifting species on the fill gas molecules, $n$ is the number density of molecules, and $f$ is an energy-exchange-efficiency factor for the 
 scattering of charge carriers from gas molecules. This condition amounts to requiring that the work done 
 by the drift field on a charge carrier between collisions and not 
 lost to collisions, must be much smaller than the carrier's thermal energy. 
 If this condition is fulfilled it will ensure that the drifting carriers' random (thermal) velocity remains consistent with 
 the bulk gas temperature. A larger scattering cross section $\sigma$ or a more 
 effective energy exchange due to strong inelastic scattering processes will lead to a 
 shorter effective mean free path and a larger value of $E_d^{max}$. Importantly, 
 $E_d^{max}$ for electrons in a given gas generally scales inversely as the pressure, as would be expected 
 from the presence of the mean free path in the ``low field" 
 condition.            
             
 If the drift field exceeds $E_d^{max}$, the energy gained from the drift field 
 becomes non-negligible. The average energy of drifting charge carriers begins to increase appreciably, 
 giving them an effective temperature $T_{eff}$ which can be orders of magnitude 
 larger than that of the bulk gas. Under these conditions, the kinetic theory arguments underlying 
 equation~\ref{eq:diff} remain approximately valid if the gas temperature $T$ is replaced by 
 $T_{eff}$. Diffusion stops dropping with increasing drift field and may rapidly {\it 
 increase} in this regime, with longitudinal diffusion increasing more rapidly than transverse. 
           
 Values of $E_d^{max}/P$ for electrons drifting in various gases and gas mixtures 
 vary from $\sim$0.1--1 V/cm/Torr at 300 K~\cite{SauliBible,Caldwell}. 
 With drift fields limited to this range and a gas pressure of $\sim$ 50 Torr, the rms diffusion for 
a 1 meter drift distance would be several mm, severely degrading the tracking resolution.

Effects of diffusion can be significantly reduced  by drifting negative ions instead of 
electrons~\cite{Martoff2000,Martoff2009,Ohnuki:NIMA2001}.  
Electronegative vapors
have been found which, when mixed into detector gases, reversibly capture primary 
ionization electrons within $\sim$ 100 $\mu$m of their creation.  The resulting negative ions drift
to the gain region of the chamber, where collisional processes free the electrons and initiate normal Townsend avalanches~\cite{Dion2009}.
Ions have E$_d^{max}$ values corresponding to E/P = 20 V/cm Torr and higher.  
This is because the ions' masses are comparable to the gas molecules, so the energy-exchange-efficiency factor $f$ which determines 
$E_d^{max}$ is much larger than for electrons. Ion-molecule scattering cross sections also tend to be larger than electron-molecule cross sections.  The use of negative ion drift in TPCs would allow sub-millimeter rms diffusion for drift distances of 1 meter or larger, although total drift voltage differences in the neighborhood of 100 kV would be required.   
         
 The above outline shows that diffusion places serious constraints on the design 
 of detectors with large sensitive mass and millimeter track resolution, particularly when 
 using a conventional electron drift TPC.

 \subsection{Challenges of Directional Detection  } 

 The current limits on spin-independent interactions of WIMPs in the 
 60 GeV/c$^2$ mass range have been set using 300-400 kg-day exposures, for 
 example by the XENON10~\cite{XENON2008} and CDMS~\cite{CDMS2009} experiments. 
 Next generation non-directional experiments 
 are being planned to achieve zero background with hundreds or thousands of times larger 
 exposures~\cite{Arisaka2009}.  

To be competitive, directional detectors should be able to use comparable 
exposures. However, integrating large exposures is particularly difficult for 
low-pressure gaseous detectors.
 A fiducial mass of a few tons will be necessary to observe DM-induced nuclear recoils 
for much of the theoretically-favored range of parameter space~\cite{Jungman1996}. 
This mass of low-pressure gas would occupy thousands of cubic meters. 
 It is, therefore, key to the success of the directional DM program to develop detectors 
with a low cost per unit volume. Since for standard gaseous detectors the largest expense is represented by the 
cost of the readout electronics, 
it follows that a low-cost read-out is essential to make DM directional detectors financially viable.

\section{Dark Matter TPC Experiments}       
         
 \subsection{Early History of Direction-Sensitive WIMP Detectors}       
 As early as 1990, Gerbier {\em et al.}~\cite{Gerbier1990} discussed using a hydrogen-filled TPC at 0.02 bar, 
drifting electrons in a 0.1 T magnetic field to detect proton recoils from Dark Matter collisions. This 
proposal was made in the context of the ``cosmion", a then-current WIMP candidate with very large 
(10$^{-36}$ cm$^2$) cross section for scattering on protons.  These authors explicitly considered the 
directional signature, but they did not publish any experimental findings.    
             
 A few years later, the UCSD group led by Masek~\cite{Buckland1994} published results 
 of early trials of the first detector system specifically designed for a 
 direction-sensitive Dark Matter search. This pioneering work used optical readout of light 
 produced in a parallel plate avalanche counter (PPAC) 
 located at the readout plane of a low-pressure TPC. 
 The minimum discernible track length was about 5 mm. 
 Electron diffusion at low pressures and its importance for the 
 performance of gas detectors was also studied~\cite{MattDiff}. 
This early work presaged some of the most recent developments in the field, described in section~\ref{DMTPC}. 
             
 \subsection{DRIFT}            
 The DRIFT-I collaboration~\cite{Snowden-Ifft:PRD2000} mounted the first underground 
experiment designed for direction sensitive 
 WIMP recoil detection~\cite{Alner2004}. Re-designed detectors were built and further 
 characterization measurements 
 were performed by the DRIFT-II~\cite{Lawson2005} collaboration. 
Both DRIFT detectors were cubical 1 m$^3$ negative-ion-drifting 
TPCs with two back-to-back 0.5 m drift spaces. To minimize material 
 possibly contributing radioactive backgrounds, the central drift cathode was 
designed as a plane of 
 20 micron wires on 2 mm pitch. The endcap MWPCs used 20 $\mu$m anode wires on 
 2 mm-pitch, read out with transient digitizers. In DRIFT-II the 
 induced signals on grid wires between the MWPC anode and the drift 
 space were also digitized. DRIFT-I had an amplifier- and digitizer-per-wire readout, while
 DRIFT-II signals were cyclically grouped onto a small number of amplifiers and 
 digitizers. Both detectors used the negative ion drift gas CS$_2$ at nominally 
 40 Torr, about one eighth of the atmospheric pressure. 
The 1 m$^3$ volume gave approximately 170 grams of target 
 mass per TPC. The CS$_2$ gas fill allowed diffusion suppression by running with very high drift fields despite the low pressure. DRIFT-II used drift fields up to 624 V/cm (16 V/cm/Torr).

 The detectors were calibrated with alpha particles, $^{55}$Fe x-rays and $^{252}$Cf neutrons. 
 Alpha particle Bragg peaks and neutron recoil events from sources were quickly 
 seen after turn-on of DRIFT-I underground in 2001. Neutron exposures gave energy 
 spectra in agreement with simulations when the energy per ion pair W was adjusted in accordance with the discussion of ionization yields given above. Simulations of DRIFT-II showed that 
 the detector and software analysis chain had about 94\% efficiency for detection of those $^{252}$Cf 
 neutron recoils producing between 1000 and 6000 primary ion pairs, 
 and a $^{60}$Co gamma-ray rejection ratio better than a few times 10$^{-6} $~\cite{drift_II_n}. 
 A study of the direction sensitivity of DRIFT-II for neutron 
 recoils~\cite{driftIIfb}  showed that a statistical signal distinguishing the beginning and end of 
 Sulfur recoil tracks (``head-tail discrimination") was available, though its energy range and statistical 
power was  limited by the 2 mm readout pitch.    
             
 At present two 1 m$^3$ DRIFT-II modules are operating underground.  
Backgrounds due   to radon daughters implanted in the internal surfaces of the detector~\cite{drift_II_n} are 
 under study and methods for their mitigation are being developed.  The absence of nonzero spin nuclides in the CS$_2$ will require a very large increase in target mass or a change of gas fill in order to detect WIMPs with this device.    

\subsection{Dark Matter Searches Using Micropattern Gas-Gain Devices} 
  It was shown above that the event rate and therefore the sensitivity of an
  optimized tracking detector improves linearly as the track length threshold gets smaller.
  In recent years there has been widespread development of gas detectors achieving very high spatial resolution by using
  micropatterned gain elements in place of wires. 
  For a recent overview of micropattern detector activity, see Ref.~\cite{pos-sens}. 
  These devices typically have 2-D arrays of individual gain
  elements on a pitch of $\sim$ 0.1 mm. Rows of elements~\cite{Black2007} or
  individual gain elements can be read out by suitable arrangements of pickup
  electrodes separate from the gain structures, or by amplifier-per-pixel electronics integrated with 
  the gain structure~\cite{medipix}. Gain-producing structures known as GEM (Gas Electron 
Multiplier~\cite{gem}) and MicroMegas (MICRO-MEsh GAseous Structure~\cite{Giomataris1996})
  have found particularly wide application.  
            
 The gas CF$_4$ also figures prominently in recent micropattern Dark Matter search proposals. 
 This gas was used for low background work in the 
 MUNU experiment~\cite{munu} and has the advantage of high $E_d^{max}$, allowing 
relatively low diffusion for electron drift at high drift field and reduced pressure~\cite{Dujmic2008-327,Christo1996,Caldwell}, 
though it does not approach negative ions in this regard. 
Containing the  odd-proton nuclide $^{19}$F is also an advantage since it 
confers  sensitivity to purely spin-coupled WIMPs~\cite{Ellis1991}, allowing smaller active mass 
experiments to be  competitive. Another attractive feature of CF$_4$ is that its 
Townsend avalanches  copiously emit visible and near infrared light~\cite{Pansky1995,Kaboth2008,Fraga2003}, 
allowing optical readout as in the DMTPC detector discussed in section~\ref{DMTPC}. 
The ultraviolet part of the spectrum may also be seen by making use of a wavelength shifter. 
Finally, CF$_4$ is non-flammable and non-toxic, and, therefore, safe to operate underground. 
             
 The NEWAGE project is a current Dark Matter search program led by a Kyoto University 
group.  This group has  recently  published the first limit on  Dark Matter interactions 
 derived from the absence of a directional modulation during a 0.15 kg-day exposure~\cite{Miuchi2007-58}. 
NEWAGE   uses CF$_4$-filled TPCs with a microwell gain 
structure~\cite{Miuchi2003,Tanimori2004,Miuchi2007-43}. 
The detector had an active volume of 23 x 28 x 30  cm$^3$  and contained CF$_4$ at 150 
Torr. Operation at higher-than-optimal gas pressure was chosen to enhance the HV 
stability of the gain structure.  The chamber was read out by a single 
 detector board referred to as a ``$\mu$-PIC", preceded by a GEM for extra gas gain. The 
 $\mu$-PIC has a micro-well gain structure produced using multi-layer printed circuit board technology.  It is  
 read out on two orthogonal, 400 micron-pitch arrays of strips. One 
 array is connected to the central anode dots of the micro-well gain structure, and the 
 other array to the surrounding cathodes.  The
 strip amplifiers and position decoding electronics are on-board with the gain structures themselves, using 
 an 8 layer   PCB structure. 
             
 The detector was calibrated with a $^{252}$Cf neutron source.  Nuclear recoils were 
detected and compared to a 
 simulation, giving a detection efficiency rising from zero at 50 keV to 
 90\% near 250 keV. For comparison, the maximum energy of a $^{19}$F recoil from 
 an infinitely heavy WIMP with the galactic escape speed is about 180 
 keV.  The measured rejection factor for $^{137}$Cs gamma rays was about 10$^{-4}$.  The 
angular resolution was reported as 25$^{\circ}$ HWHM. Measurement of the forward/backward 
sense of the tracks (``head-tail" discrimination) was not reported. 

Another gaseous Dark Matter search collaboration known as MIMAC~\cite{santos2006} is led by a group at IPN 
Grenoble, and has reported work toward an electronically 
read-out direction sensitive detector.  
They proposed the use of $^3$He mixtures with isobutane near 1 bar, and also CF$_4$ gas fills to check the dependence 
on the atomic number A 
of any candidate Dark Matter signal.  The advantages claimed for  $^3$He as a Dark 
Matter search target include  nonzero nuclear spin,  low mass and hence sensitivity to 
low WIMP 
masses, and a very low Compton cross section which suppresses backgrounds from gamma 
rays.  The characteristic (n,p) capture interaction with slow neutrons gives a strong 
signature for the presence of slow neutrons.   
The ionization efficiency of $\sim$ 1 keV $^3$He recoils is also expected to be  very high, allowing efficient detection of the small energy releases expected for this target and for light WIMPs.  A micropattern TPC with $\sim$ 350 $\mu$m anode pitch was proposed  
to obtain the desired electron rejection factor at a few keV.  
The MIMAC collaboration uses an ion source to generate monoenergetic  $^3$He and F   ions  
for measuring the ionization yield in their gas mixtures~\cite{Guillaudin:2009fp}.  

\subsection{DMTPC}
\label{DMTPC} 
The Dark Matter Time Projection Chamber (DMTPC) 
collaboration has developed a new detector concept~\cite{Sciolla:2009fb} that 
addresses the issue of scalability of directional Dark Matter detectors by using optical readout, 
a potentially very inexpensive readout solution. 

The DMTPC detector~\cite{Sciolla:2008ak,Sciolla:2008mpla} is a low-pressure TPC 
filled with  CF$_4$ at a nominal pressure of 50 torr. The detector is  read out by 
an array of  CCD cameras and  photomultipliers (PMTs) mounted outside the vessel 
to reduce the amount of radioactive material in the active volume.  
The CCD cameras image the visible and near infrared photons that are produced by the avalanche 
process in the 
amplification region, providing a  projection of the 3-D nuclear recoil 
on the 2-D amplification plane. 
The 3-D track length and direction of the recoiling nucleus is reconstructed 
by combining the measurement of the projection along the amplification 
plane (from pattern recognition in the CCD) with the projection along the 
direction of drift, determined from the waveform of the signal from the PMTs. 
The sense of the recoil track is determined by measuring dE/dx  along the length of the track. 
The correlation between the energy of the recoil, proportional to the number of photons collected in the 
CCD, and the length of the recoil track provides  an excellent rejection of all electromagnetic backgrounds. 
 
Several alternative implementations of the amplification region~\cite{Dujmic2008-58} were developed. 
In a first design, the amplification was obtained by 
applying  a large potential difference ($\Delta$V = 0.6--1.1 kV) 
between a copper plate and a conductive woven mesh kept at a uniform distance 
of 0.5 mm. 
The copper or stainless steel mesh was made of 28 $\mu$m wire with a pitch of 256 $\mu$m. 
In a second design 
the copper plate was replaced with two additional woven meshes.  
This design has the advantage of creating a transparent amplification region, 
which allows a substantial cost reduction since a single CCD camera can image tracks 
originating in two drift regions  located on either side of a single amplification 
region.

The current DMTPC prototype~\cite{dujmicICHEP} 
consists of two optically independent regions contained in one stainless steel vessel. Each region is a cylinder with 30 cm diameter and 20 cm height contained inside a field cage. 
Gas gain is obtained using the mesh-plate design described above. The detector is 
read out by two CCD cameras, each imaging one drift region. 
Two f/1.2 55 mm Nikon photographic lenses focus light onto  
two commercial Apogee U6 CCD cameras equipped with  Kodak 1001E CCD chips. 
Because the total area imaged is $16\times16$~cm$^2$, the detector has an active volume of about 10 liters.
For  WIMP-induced nuclear recoils of 50 keV, the energy and angular resolutions 
obtained with the CCD readout were estimated to be $\approx$ 15\% and 25$^{\circ}$, respectively. 
This apparatus is currently being operated above ground with the goal of characterizing the detector response and 
understanding its backgrounds. A second 10-liter module is being constructed for underground operations at 
the Waste Isolation Pilot Plant (WIPP)  in New Mexico. 

A 5.5 MeV alpha source from  $^{241}$Am is used to study the gain of the 
detector as a function of the voltage and gas pressure, as well as to 
measure the resolution as a function of the drift distance of the 
primary electrons to quantify the effect of the transverse diffusion. 
These studies~\cite{Dujmic2008-327,Caldwell} 
show that the transverse diffusion allows for a sub-millimeter spatial resolution 
in the reconstruction of the recoil track for  drift distances up to 20--25 cm. 
The gamma ray rejection factor, measured using a $^{137}$Cs source, 
is better than 2 parts per million~\cite{Dujmic2008-327}.

The performance of the DMTPC detector in determining the sense and direction 
of nuclear recoils has been evaluated by studying the recoil of fluorine nuclei 
in interaction with low-energy neutrons. 
The initial measurements were obtained running the chamber at 280 Torr and 
using 14 MeV neutrons from a deuteron-triton generator and a $^{252}$Cf  source. 
The ``head-tail'' effect was clearly observed~\cite{Dujmic2008-327,Dujmic:2008iq}  
for nuclear  recoils with energy between 200 and 800 keV. 
Better sensitivity  to lower energy thresholds was achieved by using higher gains and  
lowering the CF$_4$ pressure to 75 torr. 
These measurements demonstrated~\cite{Dujmic2008-58} ``head-tail'' discrimination for recoils above 100 keV, 
and reported a good agreement with the predictions of the SRIM~\cite{SRIM} simulation.  
``Head-tail'' discrimination is expected to extend to recoils above 50 keV when the detector is 
operated at a pressure of 50 torr. 
To evaluate the event-by-event ``head-tail'' capability of the detector as a function of the 
energy of the recoil, the DMTPC collaboration introduced a quality factor 
$Q(E_R) = \epsilon(E_R) \times (1 - 2 w(E_R))^2$, 
where $\epsilon$ is the recoil reconstruction  efficiency and $w$ is the fraction of wrong ``head-tail'' 
assignments. The $Q$ factor represents the effective 
fraction of reconstructed recoils with “head-tail” information, and the error 
on the “head-tail” asymmetry scales as $1/\sqrt(Q)$.  Early measurements demonstrated 
a $Q$ factor of 20\% at 100 keV and 80\% at 200 keV~\cite{Dujmic2008-58}.

The DMTPC collaboration is currently designing a  1-m$^3$  detector.  
The apparatus consists of a stainless steel vessel of 1.3 m diameter and 1.2 m height.
Nine CCD cameras and nine PMTs are mounted on each of the top and bottom plates of the 
vessel, separated from the active volume of the detector by an acrylic window.  
The detector consists of two optically separated regions. Each of these regions is 
equipped with a triple-mesh amplification device, located between two symmetric 
drift regions.  Each drift region has a diameter of 1.2 m and a height of 25 cm, for 
a total active volume of 1 m$^3$. 
A field cage made of stainless steel rings keeps the uniformity of the electric field within 1\% in the fiducial volume. A gas system recirculates and purifies the CF$_4$.

When operating the detector at a pressure of 50 torr,
a 1 m$^3$  module will contain 250 g of CF$_4$. 
Assuming a detector  threshold of 30 keVee (electron-equivalent energy, corresponding to 
nuclear recoil energy threshold  $\sim$ 50 keV), and 
an overall data-taking efficiency of 50\%, 
a one-year underground run will yield an exposure of 45 kg-days. 
Assuming negligible backgrounds,  such an exposure 
will allow the DMTPC collaboration to improve the current limits 
 on spin-dependent interactions on protons 
by about a factor of 50~\cite{Dujmic2008-58}.

\section{ Conclusion } 
Directional  detectors can  provide an unambiguous positive 
observation of Dark Matter particles even in presence of 
insidious backgrounds, such as neutrons or neutrinos. 
Moreover, the dynamics of the galactic Dark Matter halo will be revealed
by measuring the direction of the incoming WIMPs, 
opening the path to WIMP astronomy.  

In the past decade, several groups have investigated new ideas to develop directional 
Dark Matter detectors. Low-pressure TPCs are best suited for this purpose if 
an accurate (sub-millimeter) 3-D reconstruction of the nuclear recoil 
can be achieved. 
A good tracking resolution also allows for an effective rejection of all 
electromagnetic backgrounds, in addition to statistical discrimination against 
neutrinos and neutrons based on the directional signature. 
The choice of different gaseous targets makes these  detectors well suited for 
the study of both spin-dependent (CS$_2$) or spin-independent (CF$_4$ and $^3$He) interactions. 

A vigorous R\&D program has explored both electronic and optical readout solutions, demonstrating 
that both technologies can effectively and efficiently reconstruct 
the energy and vector direction of the nuclear recoils expected from 
Dark Matter interactions. 
The  challenge  for the field of directional Dark Matter detection 
is now to develop and deploy very sensitive and yet inexpensive readout solutions, 
which will make large directional detectors financially viable.

\section*{Acknowledgments}
The authors are grateful to D.~Dujmic and M.~Morii
for useful discussions and for proofreading the manuscript.  
G.~S. is supported by the M.I.T. Physics Department and the 
U.S. Department of Energy (contract number DE-FG02-05ER41360).  
C.~J.~M. is supported by Fermilab and Temple University. 

\section*{References} 
\bibliographystyle{unsrt}
\bibliography{all_DM}

\begin{thebibliography}{10}

\bibitem{Hinshaw2008}
G.~Hinshaw et~al.
\newblock {\em Astrophys. J. Suppl.}, 180:225, 2009.

\bibitem{LeeWeinberg}
B.~W. Lee and S.~Weinberg.
\newblock {\em Phys. Rev. Lett.}, 39(4):165--168, 1977.

\bibitem{Weinberg82}
S.~Weinberg.
\newblock {\em Phys. Rev. Lett.}, 48:1303--1306, 1982.

\bibitem{Jungman1996}
G.~Jungman, M.~Kamionkowski, and K.~Griest.
\newblock {\em Phys. Rept.}, 267:195, 1996.

\bibitem{Monroe2007}
J.~Monroe and P.~Fisher.
\newblock {\em Phys. Rev. D}, 76:033007, 2007.

\bibitem{Mei2006}
D.~Mei and A.~Hime.
\newblock {\em Phys. Rev. D}, 73:053004, 2006.

\bibitem{SmithLewin1990}
P.~F. Smith and J.~D. Lewin.
\newblock {\em Phys. Rep.}, 187(5):203--280, 1990.

\bibitem{Drukier1986}
A.~K. Drukier, K.~Freese, and D.~N. Spergel.
\newblock {\em Phys. Rev. D}, 33:3495, 1986.

\bibitem{Bernabei2008}
R.~Bernabei et~al. (DAMA~Collaboration).
\newblock {\em Eur. Phys. J. C}, 56:333, 2008.

\bibitem{Spergel}
D.~N. Spergel.
\newblock {\em Phys. Rev. D}, 37:1353, 1988.

\bibitem{Copi1999}
J.~Copi, C.J.and~Heo and L.M Krauss.
\newblock {\em Phys. Lett. B}, 461:43, 1999.

\bibitem{Vergados2003}
J.D. Vergados.
\newblock {\em Phys. Rev. D}, 67:103003, 2003.

\bibitem{Morgan2004}
B.~Morgan, A.~M. Green, and N.~J.~C. Spooner.
\newblock {\em Phys. Rev.}, D71:103507, 2005.

\bibitem{Freese2005}
K.~Freese, P.~Gondolo, and H.J. Newberg.
\newblock {\em Physical Review D}, 71:43516--1--15, 2005.

\bibitem{Alenazi2008}
M.~S. Alenazi and P.~Gondolo.
\newblock {\em Phys. Rev. D}, 77:043532, 2008.

\bibitem{Sikivie1999}
P.~Sikivie.
\newblock {\em Phys. Rev. D}, 60:063501, 1999.

\bibitem{Tkachev1997}
I.~I. Tkachev and Y.~Wang.
\newblock {\em Phys. Rev. D}, 56:1863, 1997.

\bibitem{Sikivie1995}
P.~Sikivie, I.~I. Tkachev, and Y.~Wang.
\newblock {\em Phys. Rev. Lett.}, 75:2911, 1995.

\bibitem{Green2007}
A.~M. Green and B.~Morgan.
\newblock {\em Astroparticle Physics}, 27:142--149, 2007.

\bibitem{rave}
M.~C.~Smith et~al.
\newblock {\em Mon.Not.Roy.Astron.Soc.}, 379:755, 2007.

\bibitem{Lindhard1963}
J.~Lindhard, M~Scharff, and H.~Schiott.
\newblock {\em Kgl. Danske Vadenskab. Selskab, Mat. Fys. Medd.}, 33:No. 14,
  1963.

\bibitem{Lindhard-int}
J.~Lindhard, M~Scharff, and H.~Schiott.
\newblock {\em Kgl. Danske Vadenskab. Selskab, Mat. Fys. Medd.}, 33:No. 10,
  1963.

\bibitem{White2007}
J.T.~White et~al.
\newblock {\em Nuclear Physics B (Proc. Suppl.)}, 173:144, 2007.

\bibitem{Martin2009}
C.~Martin et~al.
\newblock Abstract G10-7 submitted to APS April Meeting 2009.

\bibitem{Aprile2006}
E.~Aprile et~al.
\newblock {\em Phys. Rev. Lett.}, 97:081302, 2006.

\bibitem{SRIM}
J.~F. Ziegler.
\newblock {SRIM}- {T}he {S}topping and {R}ange of {I}ons in {M}atter.
\newblock www.srim.org.

\bibitem{BLD}
N.~Bohr, J.~Lindhard, and K.~Dan.
\newblock {\em Kgl. Danske Vadenskab. Selskab, Mat. Fys. Medd.}, 28:No. 7,
  1954.

\bibitem{Hitachi2008}
A.~Hitachi.
\newblock {\em Radiation Physics and Chemistry}, 77:1311--1317, 2008.

\bibitem{Ziegler1985}
J.F. Ziegler, J.P. Biersack, and U.~Littmark.
\newblock Pergamon Press, Oxford, England, first edition, 1985.

\bibitem{Sigmund1998}
P~Sigmund.
\newblock {\em Nucl. Inst. Meth. B}, 135:1, 1998.

\bibitem{MSTAR}
H.~Paul and A.~Schinner.
\newblock {\em At. Data Nucl. Data Tables}, 85:377, 2003.

\bibitem{Evans1953}
G.~E. Evans, P.~M. Stier, and C.~F. Barnett.
\newblock {\em Phys. Rev.}, 90:825, 1953.

\bibitem{Lassen1964}
N.~O.~Lassen et~al.
\newblock {\em Kgl. Danske Vadenskab. Selskab, Mat. Fys. Medd.}, 34:5, 1964.

\bibitem{Cano1968}
G.~L. Cano.
\newblock {\em Phys. Rev.}, 169:278, 1968.

\bibitem{SnowdenIfft2003}
D.~P. Snowden-Ifft et~al.
\newblock {\em Nuc. Inst. Meth. A}, 498:155--164, 2003.

\bibitem{SmithLewin1996}
J.~D. Lewin and P.~F. Smith.
\newblock {\em Astropart. Phys.}, 6:87--112, 1996.

\bibitem{Cano1965}
G.~L. Cano and R.~W. Dressel.
\newblock {\em Phys. Rev.}, 139:A1883, 1965.

\bibitem{Stone1957}
W.~G. Stone and L.~W. Cochrane.
\newblock {\em Phys. Rev.}, 107:702, 1957.

\bibitem{Phipps1964}
J.~A. Phipps, J.~W. Boring, and R.~A. Lowry.
\newblock {\em Phys. Rev.}, 135:A36, 1964.

\bibitem{Boring1965}
J.~W. Boring, G.~E. Strohl, and F.~R. Woods.
\newblock {\em Phys. Rev.}, 140:A1065, 1965.

\bibitem{McDonald1969}
J.~R. Mc~Donald and G~Sideneus.
\newblock {\em Phys. Lett A}, 28:543, 1969.

\bibitem{Price1993}
J.~L.~Price et~al.
\newblock {\em Phys. Rev. A}, 47:2913, 1993.

\bibitem{Bander1995}
S.~R.~Bandler et~al.
\newblock {\em Phys. Rev. Lett.}, 74:3169, 1995.

\bibitem{Martoff1996}
C.~J.~Martoff et~al.
\newblock {\em Phys. Rev. Lett.}, 76:4882, 1996.

\bibitem{Natsume2007}
M.~Natsume et~al.
\newblock {\em Nuc. Inst. Meth. A}, 575:439--43, 2007.

\bibitem{NygrenTPC}
D.~R. Nygren.
\newblock {\em PEP-0144, Proceedings of PEP Summer Study, Berkeley}, page~58.
\newblock 1975.

\bibitem{Fancher1979}
D.~Fancher et~al.
\newblock {\em Nucl. Instrum. Meth.}, 161:383, 1979.

\bibitem{Snowden-Ifft:PRD2000}
D.~P. Snowden-Ifft, C.~J. Martoff, and J.M. Burwell.
\newblock {\em Phys. Rev. D}, 61:101301, 2000.

\bibitem{Sciolla:2009fb}
G.~Sciolla et~al (DMTPC~Collaboration).
\newblock 2009.
\newblock arXiv 0903.3895 (astro-ph).

\bibitem{Miuchi2007-58}
K.~Miuchi et~al (NEWAGE~Collaboration).
\newblock {\em Phys. Lett. B}, 654:58, 2007.

\bibitem{Dujmic2008-58}
D.~Dujmic et~al (DMTPC~Collaboration).
\newblock {\em Astropart. Phys.}, 30:58, 2008.

\bibitem{RnB}
L.~Rolandi and W.~Blum.
\newblock {\em Particle Detection with Drift Chambers}.
\newblock Springer-Verlag, 1994.

\bibitem{SauliBible}
F.~Sauli.
\newblock {\em Experimental Techniques in High Energy Physics, T. Ferbel, Ed.},
  page~81.
\newblock Addison-Wesley Publishing, 1987.

\bibitem{Caldwell}
T.~Caldwell et~al (DMTPC~Collaboration).
\newblock 2009.
\newblock arXiv 0905.2549 (physics.ins-det).

\bibitem{Martoff2000}
C.~J.~Martoff et~al.
\newblock {\em Nucl. Instr. Meth. A}, 440:355--359, 2000.

\bibitem{Martoff2009}
C.~J.~Martoff et~al.
\newblock {\em Nuc. Inst. Meth. A}, 598:501--504, 2009.

\bibitem{Ohnuki:NIMA2001}
T.~Ohnuki, C.~J. Martoff, and D.~P. Snowden-Ifft.
\newblock {\em Nucl. Instr. Meth. A}, 463(1-2):142--148, 2001.

\bibitem{Dion2009}
M.~P. Dion.
\newblock Temple University Physics PhD Dissertation, May 2009 (available
  online through ProQuest Dissertations and Theses database.

\bibitem{XENON2008}
J.~Angle et~al (XENON~Collaboration).
\newblock {\em Phys. Rev. Lett.}, 100:02 1303, 2008.

\bibitem{CDMS2009}
Z.~Ahmed et~al (CDMS~Collaboration).
\newblock {\em Phys. Rev. Lett.}, 102:01 1301, 2009.

\bibitem{Arisaka2009}
K.~Arisaka et~al.
\newblock {\em Astropart. Phys.}, 31:63, 2009.

\bibitem{Gerbier1990}
G.~Gerbier.
\newblock {\em Nuclear Physics B, Proceedings Supplements}, 13:207--8, 1990.

\bibitem{Buckland1994}
K.~N.~Buckland et~al.
\newblock {\em Physical Review Letters}, 73:1067--70, 1994.

\bibitem{MattDiff}
M.~J. Lehner, K.~N. Buckland, and G.~E. Masek.
\newblock {\em Astropart. Phys.}, 8:43--50, 1997.

\bibitem{Alner2004}
G.~J.~Alner et~al (DRIFT~Collaboration).
\newblock {\em Nuc. Inst. Meth. A}, 535:644, 2004.

\bibitem{Lawson2005}
T.~B.~Lawson et~al (DRIFT-II~Collaboration).
\newblock {\em Nuc. Inst. Meth. A}, 555:173--83, 2005.

\bibitem{drift_II_n}
S.~Burgos et~al (DRIFT-II~Collaboration).
\newblock {\em Astropart. Phys.}, 28:409, 2007.

\bibitem{driftIIfb}
S.~Burgos et~al (DRIFT-II~Collaboration).
\newblock {\em Nucl. Instrum. Meth.}, A600:417, 2009.

\bibitem{pos-sens}
A.~J.~Boston et~al.
\newblock {\em Nucl. Inst. Meth. A}, 573:1--322, 2005.

\bibitem{Black2007}
J.~K.~Black et~al.
\newblock {\em Nucl. Inst. Meth.A}, 581:755, 2007.

\bibitem{medipix}
S.~R.~Amendolia et~al.
\newblock {\em Nucl. Inst. Meth.A}, 422:201, 1999.

\bibitem{gem}
F.~Sauli.
\newblock {\em Nucl. Inst. Meth.A}, 386:531, 1997.

\bibitem{Giomataris1996}
Y.~Giomataris et~al.
\newblock {\em Nucl. Inst. Meth. A}, 376:29, 1996.

\bibitem{munu}
C.~Amsler et~al.
\newblock {\em Nucl. Instrum. Meth. A}, 396:115, 1997.

\bibitem{Dujmic2008-327}
D.~Dujmic et~al (DMTPC~Collaboration).
\newblock {\em Nucl. Instrum. Meth. A}, 584:327, 2008.

\bibitem{Christo1996}
L.~G.~Christophorou et~al.
\newblock {\em J. Phys. Chem. Ref. Data}, 25:1341, 1996.

\bibitem{Ellis1991}
R.~J. Ellis and R.~A. Flores.
\newblock {\em Phys. Lett. B}, 263:259, 1991.

\bibitem{Pansky1995}
A.~Pansky et~al.
\newblock {\em Nucl. Instrum. Meth.}, A354:262--269, 1995.

\bibitem{Kaboth2008}
A.~Kaboth et~al (DMTPC~Collaboration).
\newblock {\em Nuc. Inst. Meth. A}, 592:63--72, 2008.

\bibitem{Fraga2003}
M.~M. F.~R. Fraga et~al.
\newblock {\em Nucl. Instrum. Meth.}, A504:88--92, 2003.

\bibitem{Miuchi2003}
H.~Kubo et~al (NEWAGE~Collaboration).
\newblock {\em Nucl. Instrum. Meth. A}, 513:94, 2003.

\bibitem{Tanimori2004}
T.~Tanimori et~al.
\newblock {\em Phys. Lett. B}, 578:241, 2004.

\bibitem{Miuchi2007-43}
K.~Miuchi et~al (NEWAGE~Collaboration).
\newblock {\em Nucl. Instrum. Meth. A}, 576:43, 2007.

\bibitem{santos2006}
D.~Santos et~al.
\newblock {\em J. Phys. Conf. Ser.}, 65:012012, 2007.

\bibitem{Guillaudin:2009fp}
O.~Guillaudin et~al.
\newblock 2009.
\newblock arXiv 0904.1667 (astro-ph).

\bibitem{Sciolla:2008ak}
G.~Sciolla et~al (DMTPC~Collaboration).
\newblock 2008.
\newblock arXiv 0811.2922 (astro-ph).

\bibitem{Sciolla:2008mpla}
G.~Sciolla.
\newblock 2008.
\newblock arXiv 0811.2764 (astro-ph).

\bibitem{dujmicICHEP}
D.~Dujmic et~al (DMTPC~Collaboration).
\newblock 2008.
\newblock arXiv 0810.2769 (physics.ins-det).

\bibitem{Dujmic:2008iq}
D.~Dujmic et~al (DMTPC~Collaboration).
\newblock {\em J. Phys. Conf. Ser.}, 120:042030, 2008.

\end{thebibliography}
 
\end{document}